\documentclass[conference]{IEEEtran}
\ifCLASSINFOpdf
\else
\fi

\usepackage{array}

\usepackage{graphicx}
\usepackage{epstopdf}
\usepackage{amssymb}
\usepackage{setspace}
\usepackage[tight,footnotesize]{subfigure}

\usepackage{algorithm}
\usepackage{algorithmicx}
\usepackage{setspace}
\usepackage{multirow}
\usepackage{amsmath}
\usepackage{xcolor}

\usepackage{algpseudocode}
\usepackage{subfigure}
\usepackage{float}
\usepackage{cite}
\usepackage{url}

\def\BibTeX{{\rm B\kern-.05em{\sc i\kern-.025em b}\kern-.08em T\kern-.1667em\lower.7ex\hbox{E}\kern-.125emX}}

\allowdisplaybreaks[4]
\begin{document}

\title{BPDS: A Blockchain based Privacy-Preserving Data Sharing for Electronic Medical Records}

\author{\IEEEauthorblockN{Jingwei Liu\IEEEauthorrefmark{1},
Xiaolu Li\IEEEauthorrefmark{1},
Lin Ye\IEEEauthorrefmark{2},
Hongli Zhang\IEEEauthorrefmark{2},
Xiaojiang Du\IEEEauthorrefmark{3}, and
Mohsen Guizani\IEEEauthorrefmark{4}}

\IEEEauthorblockA{\IEEEauthorrefmark{1}State Key Lab of ISN, Xidian University, Xi'an, 710071, China.\\ Email: jwliu@mail.xidian.edu.cn, lixiaolu0318@163.com}
\IEEEauthorblockA{\IEEEauthorrefmark{2}School of Computer Science and Engineering, Harbin Institute of Technology, Harbin, 150001, China.\\ Email: hityelin@hit.edu.cn, zhanghongli@hit.edu.cn}
\IEEEauthorblockA{\IEEEauthorrefmark{3}Department of Computer and Information Sciences, Temple University, Philadelphia, PA 19122, USA.\\ Email: dxj@ieee.org}
\IEEEauthorblockA{\IEEEauthorrefmark{4}Department of Electrical and Computer Engineering, University of ldaho, Mosocow, ldaho, USA.\\ Email: mguizani@ieee.org}
}

\maketitle

\begin{abstract}
Electronic medical record (EMR) is a crucial form of healthcare data, currently drawing a lot of attention. Sharing health data is considered to be a critical approach to improve the quality of healthcare service and reduce medical costs. However, EMRs are fragmented across decentralized hospitals, which hinders data sharing and puts patients' privacy at risks. To address these issues, we propose a blockchain based privacy-preserving data sharing for EMRs, called BPDS. In BPDS, the original EMRs are stored securely in the cloud and the indexes are reserved in a tamper-proof consortium blockchain. By this means, the risk of the medical data leakage could be greatly reduced, and at the same time, the indexes in blockchain ensure that the EMRs can not be modified arbitrarily. Secure data sharing can be accomplished automatically according to the predefined access permissions of patients through the smart contracts of blockchain. Besides, the joint-design of the CP-ABE-based access control mechanism and the content extraction signature scheme provides strong privacy preservation in data sharing. Security analysis shows that BPDS is a secure and effective way to realize data sharing for EMRs.
\end{abstract}
\IEEEpeerreviewmaketitle

\section{Introduction}

Electronic medical records (EMRs) are often highly sensitive private information for clinical diagnosis and treatment in healthcare. EMRs sharing is considered to be a promising approach to improve the quality of healthcare services, accelerate biomedical discoveries, and reduce medical costs~\cite{lo2016mobile, xia2017medshare}. However, most of private clinics and institutions usually use internal network to keep track of their patients but don't implement data sharing with other healthcare institutions, which leads to the difficulty and expense of medical service and the phenomenon of information island. To handle these issues of existing healthcare system, a secure data sharing infrastructure is necessary to be constructed.

However, in healthcare domain, three factors are particularly important: privacy, security, and interoperability. First, EMRs often have high privacy-sensitive, thus the leakage of these data could hurt patients' reputation and finances. Second, the existing healthcare systems are based on a centralized architecture, which has security and robustness vulnerabilities such as single-point-of-failure and arbitrary modification attacks. Moreover, the interoperability between healthcare institutions still remains a severe challenge. Fortunately, an emerging technology named blockchain provides a brand-new approach to solve these issues with decentralized architecture.

Blockchain technology is the underlying technology of Bitcoin~\cite{nakamoto2008bitcoin} that invented by mysterious Satoshi Nakamoto in 2008. Due to the attractive features, such as transparent, anonymous, autonomous, and tamper-proof, blockchain technology has been widely used in voting, supply chain, healthcare, IoT, and other applications~\cite{tama2017critical}. In the blockchain, all transactions are validated through consensus mechanism in the untrusted environment and no participants can modify the data arbitrarily. Blockchain is implemented in a decentralized network of computing nodes, which makes it robust against failures and attacks. Moreover, blockchain together with the smart contracts can enhance the interoperability of health data. Therefore, blockchain has a strong potential application in healthcare~\cite{rabah2017challenges}. Aiming at data security and patients' privacy issues in healthcare, we adopt consortium blockchain that is managed by several preselected medical institutions to construct a secure EMRs sharing system, because, compared with public blockchain, it can control user nodes in or out of the network through flexible access mechanisms with better privacy preservation. Besides, it has advantages of lower cost, higher performance and scalability.

So far, there have been lots of studies on the security and privacy issues in different application scenarios~\cite{du2008security, zhou2013prometheus, zhang2015interference, hei2013pipac, Du2009A, wu2016effective, zhang2015toward}. Recently, many companies and research institutions, such as Philips, Gem Health, Google, and IBM are actively exploring the medical applications based on blockchain technology. We summarize  some research efforts in healthcare~\cite{yue2016healthcare, azaria2016medrec, guo2018secure, biswas2014cloud} as follows. Yue et al. proposed a healthcare data gateway (HDG) based on the blockchain storage platform, which allows patients to process their own data without violating privacy~\cite{yue2016healthcare}. In~\cite{azaria2016medrec}, Azaria et al. presented a decentralized record management system, called MedRec, which can handle EMRs using blockchain technology. In~\cite{guo2018secure}, Guo et al. proposed an attribute-based signature scheme with multiple authorities. However, most studies only considered privacy preservation for access control mechanisms instead of the sensitive data itself. In fact, protecting data privacy from the perspective of data itself is simpler and more effective than the access control mechanism.

In certain scenarios, the sensitive information (eg. patients' name or ID number) in EMRs do not need to provide for further analysis and research. So we creatively adopt the content extraction signature (CES) to protect patients' privacy in terms of data itself, rather than just depending on access control mechanisms. CES allows patients to selectively share the signed medical data that can be verified by any others. In this paper, we propose a blockchain based privacy-preserving data sharing for EMRs, named BPDS. In BPDS, the original EMRs are stored in cloud and only the indexes are reserved in a tamper-proof blockchain. We design an improved delegated proof of stake consensus to provide the suitable and reasonable transaction verification. Secure data sharing can be accomplished through the smart contracts in blockchain. By implementing the proposed BPDS, patients can completely control their own EMRs and users or medical institutions can use data conveniently without leaking the patients' privacy.

The remainder of this paper is organized as follows. In section II, we briefly introduce some preliminaries. In section III, we describe system architecture and the implementation of BPDS in detail. Section IV analyzes the security of BPDS. Finally, we conclude the paper in section V.

\section{Preliminaries}
In this section, some preliminaries used in our blockchain-based data sharing scheme are introduced.

\subsection{Blockchain}

\begin{figure}[tb]

\setlength{\abovecaptionskip}{0.1cm}
\setlength{\belowcaptionskip}{-1cm}
\begin{center}
\includegraphics[height=2.8cm, width=7.8cm]{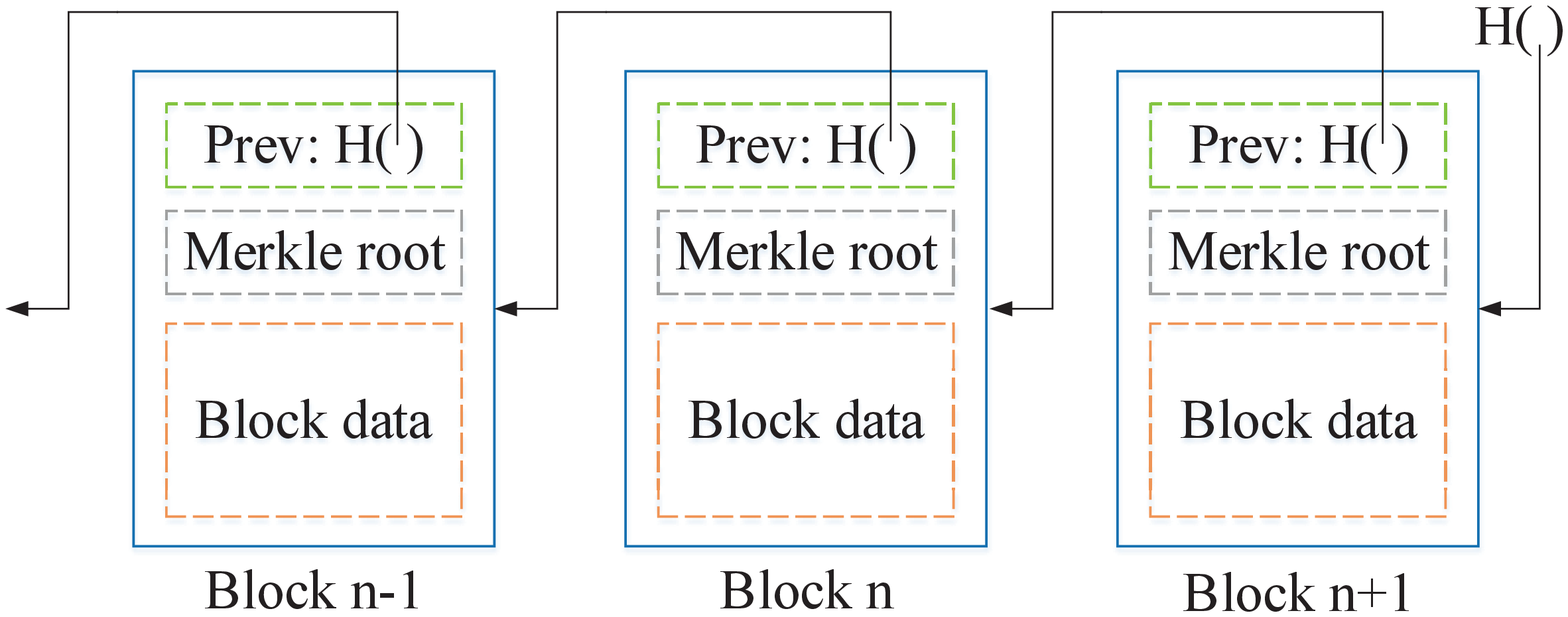}\\
\caption{A Chain of Blocks }\label{block}

\end{center}
\end{figure}

A blockchain is a type of distributed database or public ledger in which validated transactions and digital events are conserved and connected together chronologically in data blocks~\cite{zheng2017overview}, as shown in Fig.~\ref{block}. The so-called data block is composed of the data submitted by the transaction initiator and the new records produced by the transaction verifier. Moreover, each block is marked with a timestamp and the hash of the previous block, which makes the data in blockchain immutable and traceable. After reaching consensus by 51\% of the participants in the distributed network, valid blocks will be added to the blockchain.
Moreover, each node in this distributed P2P network reserves the same copy of transaction records, which provides the robustness against single-point-of-failure and attacks. Therefore, blockchain has drawn a lot of attention in various fields.

\subsection{Improved DPoS}
Delegated proof of stake (DPoS) is the backbone of BitShares. All nodes on the blockchain need to select 101 delegates. The selected 101 delegates are responsible for in turn creating validate blocks as assigned. Compared to proof of work (PoW) and proof of stake (PoS), DPoS is known as faster, more decentralized and power-saving consensus mechanism. In this paper, we use DPoS to reach consensus for each transaction in the blockchain network.

However, the original election method of DPoS cannot guarantee that the selected medical institutions are reliable. Therefore, we improve its initialization manner and elect nodes according to the rank of the medical institutions' credit scores. The top 30 institutions are designated as the representative nodes ($RPNs$), in turn, to create blocks. The next 20 institutions are designated as the audit nodes ($ATNs$) to audit these blocks. Any node that contributes to the healthcare data sharing system will obtain the corresponding reward with credit scores. If the $RPN$ misses signing the assigned block or the $ATN$ makes a incorrect audit, their credit scores will be reduced. Once the total scores fall below the threshold, this node will be replaced by the other node with higher scores.

\subsection{Smart Contracts}
Smart contracts are event-driven computer programs running on the public ledger. It can handle and transfer assets of considerable value. A famous application of smart contracts is Ethereum that is an open source blockchain platform~\cite{wood2014ethereum}. Specifically, smart contracts are some scripts or codes that are deployed in blockchain. Once the predefined conditions are activated, the scripts on the contract content could be executed without the help of an external trusted authority. The entire process is automated and the executed transactions are recorded in the public ledge for auditing. The asset owner has the right to revoke the access permissions to the user who violates the contract. In the proposed BPDS, patients are allowed to predefine access permissions, access actions (read, write, or copy) and duration in the smart contracts to finely control the data sharing of EMRs.

\subsection{Content Extraction Signature}

The content extraction signature (CES) first proposed by Steinfeld et al. in~\cite{steinfeld2001content} allows the users to remove sensitive portions from the original signed message and regenerate valid extraction signatures by themselves without extra interactions. In addition, it has the merits of low communication overhead, high efficiency and privacy preservation. So, different CES schemes have been widely used in e-commerce, e-governance, smart grid, healthcare and so on.

\begin{figure}[tb]
\setlength{\abovecaptionskip}{0.1cm}
\setlength{\belowcaptionskip}{-1cm}
\begin{center}
\includegraphics[height=6.5cm, width=8.5cm]{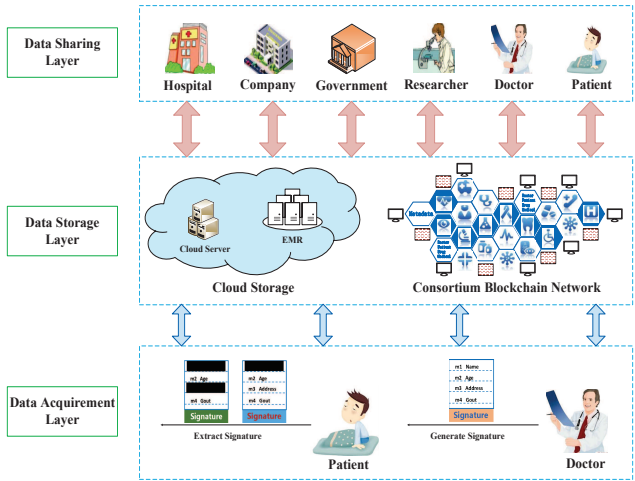}\\
\caption{System architecture of BPDS}\label{system}
\end{center}
\end{figure}

\section{BPDS: Blockchain based privacy-preserving Data sharing for EMRs}
In this section, we propose a privacy-preserving data sharing based on blockchian, called BPDS.
\subsection{System Architecture}
BPDS is devised in a three-layer architecture, consisting of data acquirement layer, data storage layer and data sharing layer, as shown in Fig.~\ref{system}. The function of each layer is described as follows.

\begin{itemize}
 \item \textbf{Data Acquirement Layer}. In this layer, EMRs are created by data providers such as doctors. Doctors sign patients' EMRs using a CES scheme and send them to the patients. Patients are the owners of EMRs and can completely control them. In order to avoid privacy information is leaked in the process of data sharing, patients can remove sensitive information of EMRs and generate valid extraction signatures.

 \item \textbf{Data Storage Layer}. The function of this layer is to store the original EMRs and its indexes. Components of data storage layer include:
 \begin{itemize}
   \item\textbf{Cloud Storage}. The cloud stores patients' encrypted EMRs and the extraction signature, meanwhile, outputs the storage location $url$ and a timestamp. Data access records also should be preserved to track the malicious entity when data leakage takes place.

    \item\textbf{Consortium Blockchain Network}. We use consortium blockchain to reserve indexes of EMRs and achieve data sharing. The patient predefined access permissions in the smart contracts to ensure data sharing securely. Besides, each access request and access activity should be recorded in the blockchain network for future auditing or investigation.
  \end{itemize}
 \item \textbf{Data Sharing Layer}. In this layer, the authorized patients, medical workers and healthcare institutions can request patients' EMRs and utilize them for making personal health plans, getting better clinic treatment or carrying out medical research.

\end{itemize}

\subsection{Design Objectives}

Because blockchain can manage medical data transparently and securely, it has attracted a lot of attention in the healthcare. The proposed BPDS aims to achieve secure storage and sharing for EMRs through the joint-design of the consortium blockchain, the cloud storage, and the context extraction signature. Meanwhile, it provides the following privacy:

\begin{itemize}
  \item Patients participate in the EMRs sharing transactions voluntarily and anonymously;
  \item Patients and data users register unique and non-identity accounts in the cloud database;
  \item The indexes of EMRs reserved in the consortium blockchain cannot be changed by any opponents;
  \item Malicious entities can be tracked when data leakage is detected;
  \item Patients can define (add/remove) who are allowed to access medical data through smart contracts. Only the authorized users can access the specified data.

\end{itemize}
\subsection{Implementation of BPDS}
In this section, we describe the workflow of the BPDS system in detail. A patient $(P)$ goes to visit his/her doctor $(D)$ and the doctor $(D)$ integrates related medical data as the EMRs for the patient $(P)$. Upon receiving the EMRs, $P$ stores them to the cloud and submits the indexes of the EMRs to the consortium blockchain with the list of authorized data users $(U)$. BPDS allows patients to manage their own EMRs as their wish. Based on the blockchain, it achieves privacy-preserving EMRs sharing through the following phases.

\begin{table}
\setlength{\abovecaptionskip}{0.1cm}
\setlength{\belowcaptionskip}{-1cm}
  \centering
  \caption{Notations} \label{notations}
   \scriptsize
  \tabcolsep 0.02in
  \setlength{\extrarowheight}{0.15cm}
 \begin{tabular}{l|l|l|l}
    \hline
     \raisebox{0.1cm}{Notations}   & \raisebox{0.1cm}{Description} & \raisebox{0.1cm}{Notations}   & \raisebox{0.1cm}{Description}  \\
    \hline
     \raisebox{0.1cm}{$P$}  &  \raisebox{0.1cm}{Patient}  & \raisebox{0.1cm}{$M/M'$}      &  \raisebox{0.1cm}{Medical data/the extracted subdata}   \\
     \raisebox{0.1cm}{$D$}  &  \raisebox{0.1cm}{Doctor}   & \raisebox{0.1cm}{$Index_i$}  &  \raisebox{0.1cm}{The indexes of EMRs}    \\
     \raisebox{0.1cm}{$U$}  &  \raisebox{0.1cm}{Data user}& \raisebox{0.1cm}{$CEAS$}      &  \raisebox{0.1cm}{Content extraction access structure} \\
     \raisebox{0.1cm}{$RPN$}  &  \raisebox{0.1cm}{The representative node}  & \raisebox{0.1cm}{$CI(M')$} &  \raisebox{0.1cm}{The extraction subset}   \\
     \raisebox{0.1cm}{$ATN$}  &  \raisebox{0.1cm}{The auditing node}        & \raisebox{0.1cm}{$T$}      &  \raisebox{0.1cm}{A tag of CES}   \\
     \raisebox{0.1cm}{$PK,SK$}  &  \raisebox{0.1cm}{$D$'s key pair for CES} & \raisebox{0.1cm}{$A$}      &  \raisebox{0.1cm}{Access Control Policy of CP-ABE}\\
     \raisebox{0.1cm}{$K_{doc}$}  &  \raisebox{0.1cm}{$D$'s encryption key} & \raisebox{0.1cm}{$t$}      &  \raisebox{0.1cm}{A timestamp}  \\
     \raisebox{0.1cm}{$pk_i,sk_i$}  &  \raisebox{0.1cm}{Key pairs of nodes} & \raisebox{0.1cm}{$H$}      &  \raisebox{0.1cm}{A hash function}\\
    \hline
 \end{tabular}
\end{table}

\subsubsection{System Setup}
To implement BPDS, users should register unique accounts and create their keys at the first. Each doctor $D$ uses a pair of keys ($PK,SK$) to generate the content extraction signature on EMRs for authentication. $D$ also needs a symmetric key $K_{doc}$ to encrypt EMRs for confidentiality. In cloud storage, the cloud server publishes keys based on CP-ABE for secure storage. All participants in the blockchain have key pairs ($pk_i,sk_i$) to complete data sharing transactions. The notations used in this paper are given in TABLE I.

\subsubsection{Data Acquiring}

In BPDS, we use the CES scheme in~\cite{Wang2015An} that can remove sensitive information from the original message to protect $P$s' privacy. $D$ divides EMRs into seven parts (Name, Gender, Age, ID number, medical history, physical examination or laboratory test, medical prescription) that are denoted as $M=\{m_1,m_2,m_3,m_4 ,m_5 ,m_6 ,m_7\}$. Then, $D$ defines the content extraction access structure $CEAS=\{2,3,5\}$ to prevent malicious extraction. $D$ selects a CES-Tag randomly with a fixed length of 80 bits, defined as $T$. The process of CES is as follows:

\begin{itemize}
\item [-]\textbf{Key Generation $(KG)$}: The certification authority chooses a a large prime $p$, a generator $g$ in $Z_p$, and a hash function: $H$: $\{0,1\}^* \rightarrow Z_p$. Then, $D$ selects a random number $a \in Z_p^*$ and calculates $v=g^a (mod$ $p)$. $D$ publishes $PK=\{p,g,v\}$ as the public key and keeps $SK=a$ as the private key.

\item [-]\textbf{Signature Generation $(SIG)$}: After generating all the keys, $D$ signs the original medical data $M$ by the Algorithm 1 as follows.
\end{itemize}

\begin{algorithm}[h]

\caption{$SIG(SK,M,CEAS)$} 
\hspace*{0.02in} {\bf Input:} 
$D$'s private key, $SK=a$; $P$'s EMRs, $M$; the content extraction access structure, $CEAS$.\\
\hspace*{0.02in} {\bf Output:} 
The full signature result.
\begin{algorithmic}
\State Select a random $k\in Z_{p-1}^*$, compute $r= g^{k}(mod$ $p)$; 
\For{each $i\in [1,7]$} 
　　\quad \State \quad $h_i\leftarrow H(M_i\|CEAS\|T\|i)$;
\EndFor
\For{each $h_i, i\in [1,7]$} 
    \State  $\delta_i\leftarrow(h_i-a\cdot r)\cdot k^{-1} mod(p-1)$;　
\EndFor\\
\Return $\delta_{Full}\leftarrow(CEAS\|T\|\delta_1\|\delta_2\|\cdots\|\delta_7)$;
\end{algorithmic}
\end{algorithm}

When the signature algorithm is completed, $D$ encrypts message $(M\|h_{i(i\in [1,7])}\|\delta_{Full}\|CEAS\|T)$ with $K_{doc}$ and encrypts his/her symmetric encryption key with $P$'s public key $pk_{pat}$. Then, $D$  sends both two encrypted information to $P$:

\begin{equation}
\setlength{\abovedisplayskip}{-1pt}
\setlength{\belowdisplayskip}{8pt}
\begin{split}
  Info=&\{E_{K_{doc}}(M\|h_{i(i\in [1,7])}\|\delta_{Full}\|CEAS\|T), \\
         &
          E_{pk_{pat}}(K_{doc})\}
\end{split}
\end{equation}

\subsubsection{Data Storing}
After receiving the encrypted information from $D$, $P$ decrypts $K_{doc}$ and further obtains $M$. Next, $P$ verifies the correctness of the full signature $\delta_{Full}$ with two steps:

\begin{itemize}
	\item [-]For each subdata $M_i$ of data $M$, compute $h_i=H(M_i \|CEAS\|T\|i)$, where $1\leq i\leq7$;

	\item [-]Extract $\delta_i$ from the full signature $\delta_{Full}$, verify $v^r\cdot r^{\delta_i}=g^{h_i} (mod$ $p)$ is hold or not, that is, $v^r\cdot r^{\delta_i}=g^{a\cdot r}\cdot r^{\delta_i }=g^{a\cdot r} \cdot g^{k\cdot {\delta_i}}=g^{a\cdot r+k\cdot {\delta_i} }=g^{h_i}$.
\end{itemize}

If the signature $\delta_{Full}$ is valid, perform the following step; Otherwise, it returns failure.

\textbf{Signature Extraction}: $P$ can extract the signature according to $CEAS$ and his/her wishes, as shown in Algorithm 2.

\begin{algorithm}[h]
\caption{$Ext(PK=\{p,g,v\},M,CEAS,\delta_{Full})$} 
\hspace*{0.02in} {\bf Input:} 
$D$'s public key, $PK_{doc}$; $P$'s EMRs, $M$; the content extraction access structure, $CEAS$; the full signature, $\delta_{Full}$.\\
\hspace*{0.02in} {\bf Output:} 
The extraction signature result.
\begin{algorithmic}
\State Construct extraction subset $CI(M')$ based on $CEAS$ and generate subdata $M'=\{M_i\mid i\in CI(M')\}$; 
\For{each $i\in CI(M')$} 
　　\State Extract $\delta_i$ from
 $\delta_{Full}$;
\EndFor
\For{each $i,j\in CI(M')$} 
    \State $\delta_{i1}\leftarrow\delta_1, \cdots, \delta_{ij}\leftarrow\delta_j$, where $\delta_{ij} (j\in [1,f])$; 　
\EndFor\\
\Return $\delta_{Ext}\leftarrow(CEAS\|CI(M')\|T\|\delta_{i1} \|\delta_{i2}\|\cdots\|\delta_{if})$;
\end{algorithmic}
\end{algorithm}

 After generating the extraction signature, $P$ encrypts the extraction signature and the corresponding EMRs. Then, $P$ stores them in the cloud through CP-ABE based cryptographic access control (CCAC)~\cite{cheng2012attributes} as shown in Algorithm 3.

\begin{algorithm}[htb]
\caption{ Data Storing Process in $CCAC$}
\label{alg:Framwork}
\setstretch{1.15}
\begin{algorithmic}[1]
\Require
  Data object, $(M_i\|h_i\|T)$;
  The public parameters, $PK$;
  The access control policy, $A$.
\Ensure
Data storage location, $urls$.
\State Generate a random document key, $k_i$;
\label{code:fram:generate}
\State Run the symmetric encryption algorithm $E$ to encrypt $(M_i\|h_i\|T)$ with $k_i$ and obtain the cipher-text $E_{k_i}(M_i\|h_i\|T)$, where ${i\in CI(M')}$;
\label{code:fram:run}
\State Run the encrypt algorithm $E'$ of CP-ABE to encrypt $k_i$ with $A$ and obtain the cipher-text $E'_A(k_i)$;
\State Upload the triples \{$E_{k_i}(M_i\|h_i\|T)$, $E'_A(k_i)$, $\delta_{Ext}$\} to the cloud storage and return the storage location $url_i$;
\label{code:fram:classify}
\end{algorithmic}
\end{algorithm}

Thus, the original data stored in the cloud is:

\begin{equation}\label{2}
\setlength{\abovedisplayskip}{3pt}
\setlength{\belowdisplayskip}{3pt}
Data=\{E_{k_i}(M_i\|h_i\|T),E'_A(k_i),\delta_{Ext} \}
\end{equation}

\subsubsection{Data Release}

In this phase, $P$ participates EMRs sharing transactions voluntarily and anonymously. $P$ signs the indexes of EMRs and obtains the signature $SIG_{sk_{pat}}(Index_i)$. Then, a transaction request ($Req$) is submitted to the consortium blockchain, where ${i\in CI(M')}$ and $t$ is a timestamp:

\begin{equation}
\setlength{\abovedisplayskip}{-1pt}
\setlength{\belowdisplayskip}{4pt}
\begin{split}
  Req= &\{E_{pk_{pat}}(Index_i) \|H(Index_i)\|\\
 &
SIG_{sk_{pat}}(Index_i)\|t\}
\end{split}
\end{equation}

\begin{equation}
\setlength{\abovedisplayskip}{-1pt}
\setlength{\belowdisplayskip}{4pt}
\begin{split}
Index_i=(url_i\|h_i\|t)
\end{split}
\end{equation}

After receiving the transaction request, the representative node $RPN$ is responsible for creating the assigned block. The specific consensus process using improved DPoS is described as follows:

 \begin{itemize}
 \item Step 1: $RPN$ verifies each transaction and integrates all valid data collected during the period into a data set (expressed as $D_{set}=\{Req\|t\}$). The data set, $RPN$'s digital signature and the hash of the data set compose a new data block. Then, $RPN$ broadcasts the transaction record ($Rec$) to the auditing nodes $ATNs$ for approval:

    \begin{equation}
    \setlength{\abovedisplayskip}{-1pt}
    \setlength{\belowdisplayskip}{4pt}
    \begin{split}
     RPN \rightarrow ATNs:
     Rec=&\{D_{set}\|D_{hash}\|t\|\\
     &SIG_{sk_{rpn}}(D_{set}\|D_{hash})\}
    \end{split}
    \end{equation}

    \begin{equation}
    \setlength{\abovedisplayskip}{-1pt}
    \setlength{\belowdisplayskip}{4pt}
    \begin{split}
     D_{hash}=H(D_{set}\|t)
    \end{split}
    \end{equation}

  \item Step 2: $ATNs$ verify the validity of the data block and return a reply ($Rep$) to $RPN$ that contains its audit result ($Res$) and signature:

    \begin{equation}
     \setlength{\abovedisplayskip}{-1pt}
    \setlength{\belowdisplayskip}{8pt}
    \begin{split}
    ATNs \rightarrow RPN:
    Rep=&E_{pk_{rpn}} \{(Res\|t)\| \\
     &
    SIG_{sk_{atn}}(Res\|t)\|t\}
    \end{split}
    \end{equation}

   \item Step 3: If 51\% $ATNs$ approve, it means that the new block is successfully created. $RPN$ broadcasts the data block together with $ATNs$' public keys and signatures, as shown in equation (8). All nodes on the consortium blockchain must update their data. It takes 10s for each $RPN$ to create a block. A full cycle takes about 300s ($30\ast10$), about 5 minutes. At the end of each cycle, the top 30 $RPNs$ have to readjust once.

    \begin{equation}
     \setlength{\abovedisplayskip}{-1pt}
    \setlength{\belowdisplayskip}{8pt}
    \begin{split}
    RPN \rightarrow All:
    D_{block}=&\{D_{set} \|D_{hash}\|pk_{atn_i}\| \\
     &
    SIG_{sk_{atn_i}}(Res\|t)\|t\}
    \end{split}
    \end{equation}

 \end{itemize}

\subsubsection{Data Sharing}
For secure EMRs sharing, $P$ pre-sets access permissions in the smart contracts, such as access rights, access actions (eg. read, write or copy), duration, etc. Once meeting the access condition, the smart contract is triggered automatically to execute the corresponding operation, which can ensure the legality and fairness of data sharing. EMRs sharing is completed by the following two parts:

a) Blockchain Access Authentication
 \begin{itemize}
 \item Step 1: Data Access Request: The data user $U$ initiates a EMRs sharing request transaction ($Req$) to the blockchain network. The request should include information such as the access target ($ID$), the access object ($obj$) and access content. $RPN$ receives the transaction request and checks the identity of $U$. Only $U$ is legal, the transaction data will be recorded in the blockchain.
   \begin{equation}\label{9}
    U \rightarrow RPN : Req=(ID\|obj\|i\|t),i\in[1,7]
    \end{equation}
    Annotation: Here $i$ indicates the index of the medical data content that the user $U$ wants to access.
 \item Step 2: Smart Contract Execution: If $Req$ meets access conditions, the smart contract is triggered to decrypt the indexes of EMRs with $sk_{pat}$ and return the cipher-text message of the indexes to $U$; Otherwise, the sharing request is denied.
   \begin{equation}\label{10}
   Message=E_{pk_{user}}(Index_i\|t)
   \end{equation}
\item  Step 3: Data Storage Location Extraction: $U$ decrypts the cipher-text message and obtains $Index_i$ that contains the storage location $url_i$.

 \end{itemize}
 b)	Cloud Storage EMR Sharing

With $url_i$, the user $U$ can retrieve the data object in the cloud, as shown in Algorithm 4.

\begin{algorithm}[h]
\caption{Data Retrieval Process in $CCAC$} 
\hspace*{0.02in} {\bf Input:} 
The data storage location in the cloud, $url_i$;
User private key, $SK$.\\
\hspace*{0.02in} {\bf Output:} 
Data object, $M_i$.
\begin{algorithmic}[1]
\State Retrieve $E_{k_i}(M_i\|h_i\|T)$, $E'_A(k_i)$ by $urls_i$; 
　　\If{the attribute set $S$ corresponding to SK does not satisfy $E'_A(k_i)$ implicit access control policy $A$} 
　　　　\State return failure
　　\EndIf
\State Run the Decrypt Algorithm of CP-ABE to decrypt $E'_A(k_i)$ with $SK$ and obtain $k_i$;
\State Run the symmetric algorithm to decrypt $E_{k_i}(M_i\|h_i\|T)$ with $k_i$ and obtain $M_i$;
\end{algorithmic}
\end{algorithm}

Then, $U$ should verify the signature $\delta_{Ext}$ to ensure the validity and integrity of $M'$ through the following two steps:

 \textbf{Signature Verification}:
 \begin{itemize}
 \item[-] Verify if $CEAS\subset CI(M')$. If it does, perform the following step. Otherwise, it aborts.
 \item[-] For each $i\in CI(M')$, compute $h_i=H(M_i \|CEAS\|T\|i)$ and verify $v^r\cdot r^{\delta_i}=g^{h_i} (mod$ $p)$ holds or not.
\end{itemize}

 If the extraction signature is correct, the user can perform his/her access action. Otherwise, the user can inform the cloud storage manager that the data might has been modified.

\section{Security Analysis}
In this part, we analyze the security of the proposed BPDS in terms of tamper-proof, privacy preservation, data secure storage and sharing.

\subsection{Tamper-Proof}

In BPDS, EMRs are immutable and cannot be modified arbitrarily. Since each data block contains a current timestamp and a hash of the previous block, chronologically nested blocks guarantee transactions cannot be changed unless someone can take over 51\% of the whole network computational power simultaneously. Moreover, each access request and access activity is recorded in the blockchain, any change to the data can be audited and tracked. So, the proposed BPDS can ensure tamper-proof property.

\subsection{Privacy Preservation}
As EMRs are highly sensitive private data of patients, they do not want to be disclosed without permission. In BPDS, the privacy property is ensured thanks to the following festures:

-{}- \textbf{Anonymity}. Each participant generates a unique account with a random public key. Therefore, each transaction on the blockchain is anonymous. In addition, users use different public keys for different transactions, which makes multiple transactions requested by the same user cannot be linked.

-{}- \textbf{Cloud Storage}. The original EMRs are encrypted and stored in the cloud storage. In this way, not only the problem of limited storage capacity of blockcahin is solved, but also the risk of the original medical data leakage is greatly reduced.

-{}- \textbf{Content Extraction Signature}. The proposed scheme employs CES scheme when the doctors sign the EMRs. The patients can remove any sensitive portions in the original data to obtain the valid extraction signatures with minimal risk of data privacy leakage. Moreover, any entities cannot forge extraction signatures without the signer's private key.

-{}- \textbf{Improved DPoS}. BPDS uses the improved DPoS consensus to realize the trust between a certain number of preselected nodes in the consortium blockchain. In the improved DPoS, the selected medical organizations are reputable and reliable, which guarantees the reliability of data sharing.

\subsection{Data Secure Storing and Sharing}
The security of data storing and sharing is an important feature of BPDS. In this scheme, patients can have complete control over their own EMRs. The processes from data acquiring to data sharing are all secure.

-{}- \textbf{Data Acquiring}. The use of symmetric encryption technology guarantees the confidentiality and integrity of EMRs generated by doctors.

-{}- \textbf{Data Storing}. The patient encrypts the original EMRs and stores them in the cloud. The use of the distributed storage and CP-ABE-based access control scheme in cloud ensures the security of the medical data.

-{}- \textbf{Data Release}. First, the indexes of EMRs are reserved in a tamper-proof blockchain, which cannot be modified arbitrarily. Second, blockchain is a distributed database without single-point-of-failure and each node has a copy of transaction records. Besides, the digital signature provides authentication, integrity, and non-repudiation for each transaction.

-{}- \textbf{Data Sharing}. In BPDS, the data access permissions are preset in the smart contracts. Only authorized users or institutions can use the EMRs. The executed access records are stored in the blockchain to trace the behaviours of data. Once someone violates the access rules or permissions, the data owner has the right to revoke his/her access permission.

\section{Conclusion}
In this paper, a blockchain-based privacy-preserving data sharing system for EMRs is proposed, named BPDS. In BPDS, EMRs are stored in the cloud and the indexes are recorded in a tamper-proof consortium blockchain, which solves the potential security risks of data centralized storage. The joint-design of the CP-ABE-based access control mechanism and the content extraction signature scheme provides strong privacy preservation in data sharing. Moreover, the use of smart contracts for presetting access permissions ensures data access securely. By implementing the proposed BPDS, patients can have complete control over their own EMRs and the users or institutions can use data conveniently without any risk on patients' privacy.

\section*{Acknowledgment}
This work is supported by the Key Program of NSFC-Tongyong Union Foundation under Grant U1636209, the 111 Project (B08038) and Collaborative Innovation Center of Information Sensing and Understanding at Xidian University.

\ifCLASSOPTIONcaptionsoff
  \newpage
\fi

\bibliographystyle{IEEEtran}
\bibliography{ms}

\end{document}